\documentclass[11pt,twoside]{atmp}

\usepackage{amsmath,amssymb}

\usepackage[all]{xy}
\usepackage{color}
\numberwithin{equation}{section}

\copyrightnotice{2007}{1}{49}{70} %% year, volume, first page, last page.

\begin{document}

\title[A New Formulation of General
Relativity] {A New Formulation of General Relativity - Part III:
GR as Scalar Field Theory}

\arxurl{hepreference}

\author[ Joachim Schr\"oter]{ Joachim Schr\"oter}

\address{Department of Physics, University of Paderborn,\\ D-33095 Paderborn, Germany}  %lines should be separated with double backslashes: \\
\addressemail{J@schroe.de}

\begin{abstract}
{\small{The aim of this paper (Part III) is formulating GR as a
scalar field theory. The basic structural elements of it are a
generating function, a generalized density and a generalized
temperature. One of the axioms of this theory is a generalized
Einstein equation which determines the generating function
directly. It is shown that basic concepts like orientation, time
orientation and isometry are expressible in terms of generating
functions. At the end of the paper six problems are formulated
which are still unsolved and can act as a stimulant for further
research.}}
\end{abstract}

\maketitle

\section{Heuristic Considerations}

{\bf 9.1:} In Chapter 5 of Part I (cf.\ e.g.\ Remark (5.4),
Propositon (5.15)) we have seen that the function $\Psi$ (cf.
Definition (3.7)) within the frame theory $\Phi_R$ (or
$\Phi^\ast_R$) (cf. Notation (2.1) and Chapter 6.) generates the
metric $g$ and the velocity $v$. It turned out (cf.\  Proposition
(7.1)) that $\Psi$ is a generating function in the sense of
Definition (7.2).

In this paper (the last part of a series of three papers) I want
to establish a frame theory $\Phi_{sc}$ which has only two base
sets, the set of events $M$ and the reals $\mathbb{R}$, and three
structural terms $\Psi, \tilde{\eta}$ and $\tilde{\vartheta}$,
where the two latter terms are generalizations of
 the density and the temperature. Hence, the class of systems to be considered is
 the same as in Part I.

\cutpage

The {\bf problem} to be solved in this section is this: find
axioms governing the base sets and the structural terms such that
there are reasonable models of the theory $\Phi_{sc}$.\\
The method applied for this purpose is a heuristic argumentation.
Clearly, such reasoning is not logically compelling, i.e. the
axioms can not be deduced in the strict sense.

{\bf 9.2:} In a first step the geometrical and kinematical axioms
of $\Phi_{sc}$ are considered, i.e. all those axioms which solely
refer to $\Psi$. Since $\Psi$ is intended to be a generating
function it should have the properties P1 to P5 of Section 7.1.2.
But in our case these conditions cannot serve as axioms for $\Psi$
because they contain a metric $g$ and a velocity $v$ besides
$\Psi$.

Nevertheless, let us consider that part of P1 to P5 which refers
only to $\Psi$. The result are the following three conditions Q1
to Q3:

{\bf Q 1:} $\Psi$ is a function, $\Psi: \cup_{q \in M} V_q \times
\{q\} \rightarrow \mathbb{R}^4$, where $V_q$ is a subset of $M$
and $q \in V_q$, such that $\mathcal{A}=\{(V_q, \Psi (\cdot, q)):
q \in M\}$ is a $ C^{k}$-atlas, $k \ge 3$, on $M$ and such that
$(M, \mathcal{A})$ is a connected Hausdorff manifold; the function $ \Psi$ is of class $C^{k}, k \ge 3 $.\\

{\bf Q 2:} For each $q \in M$ define $ \gamma_{q} $ by

\begin{equation} \label{Q2}
\gamma_q (t) = \Psi (\cdot,q)^{-1} (0, 0, 0, t)
\end{equation}

for all $(0, 0, 0, t) \in  \; \mathrm{ran} \; \Psi (\cdot, q)$.
Then $\mathrm{dom} \gamma_q =: J_q$, is an interval and there is a
$t_q \in {J}_q$ such that $\gamma_{q} (t_{q}) = q$.\\

{\bf Q 3:} For all $q' \in W_q := \mathrm{ran} \gamma_{q}$ the equation $\Psi (\cdot, q') = \Psi (\cdot, q)$ holds.\\

Now these conditions determine $\Psi$ to the following extent:\\
{\bf Proposition (9.1):} If $\Psi$ satisfies Condition Q1 then
there is exactly one metric $g$ and one velocity field $v$ such
that $\Psi$ is a generating function satisfying the conditions P1
to P3. If, in addition, $\Psi$ satisfies Q2 and Q3 then $\Psi$
fulfils also P4 and P5.\\

{\bf Proof:} First of all, because of Q1 Condition P1 is satisfied
for ${\mathcal{A}}^+ = {\mathcal{A}}$. Next define

\begin{equation} \label{proof9.1}
\Theta^\alpha (p) := \left. d_p \Psi^\alpha (p, q) \right|_{q=p}
\quad \text{and}  \quad e_\beta (p) := \left.
\partial_{{\Psi^\beta}(p, q)} \right|_{q=p}  .
\end{equation}

Then, in order that $\Psi$ is a generating function, the metric
$g$ and the velocity $v$ have to be given uniquely by

\begin{equation} \label{}
g = \eta_{\alpha \beta} \Theta^\alpha \otimes \Theta^\beta \quad
\text{and}  \quad  v = e_4.
\end{equation}

Hence the Conditions P2 and P3 are fulfilled. From Q2 it follows
that $\Psi^\alpha (\cdot, q) \circ \gamma_q (t) = t
\delta^\alpha_4$. Hence $\dot{\gamma}_q (t) =
\partial_{{\Psi^4}(\cdot, q)} |_{{\gamma_q}(t)}$. Using Q3 we have
$\Psi (\cdot, q) = \Psi (\cdot, \gamma_q (t))$ so that
$\dot{\gamma}_q (t) =  e_4 (\gamma_q (t))$. Hence, for each $q$
the path $\gamma_q$ is a solution of the differential equation
$\dot{\gamma}_q = v (\gamma_q)$ which has unique solutions because
$v$ is of class $C^r, r \ge 2$. Since for each
$q \in M$ we have $\gamma_q (t_q) = q$ all integral curves are of the form \ref{Q2}. This means that also P4 and P5 are satisfied.\\

This result suggests that the axioms governing $\Psi$ we are looking for are the Conditions Q1 to Q3 or some equivalents of them.\\

{\bf 9.3:} In order to complete the axioms of the theory
$\Phi_{sc}$ one has to set up equations which determine the fields
$\Psi, \tilde{\eta}$ and $\tilde{\vartheta}$ where up to now we
only know that $\tilde{\eta}$ and $\tilde{\vartheta}$ must have
something to do  with density and temperature. Clearly, the
starting point for our heuristic search are the axioms EM and EE
of
 Chapter 4 and 6 in Part I. At the same time it is clear that the equations of motion and  Einstein's equation written in terms
of $g, v, \eta$ and $\vartheta$ are not suitable to determine
$\Psi$ directly. But, since $g$ and $v$ are generated by $\Psi$
or, more precisely, since they can be expressed in terms of the
tetrads $\Theta^\alpha$ and $e_\beta, \alpha, \beta = 1, \cdots,
4$, the equations of motion and the Einstein equation are also
expressible in these terms. Thus, the problem of determining
$\Psi$ can be split up into two parts: first solve these equations
for $\Theta^\alpha$, $\eta$ and $\vartheta$, and then determine
$\Psi$ from the equations $d_p \Psi^\alpha (\cdot, q) |_{p=q} =
\Theta^\alpha (p)$, e.g. via the methods developed in Chapter 8.
Such procedure is possible. But the theory $\Phi_{vs}$ thus
obtained is not a scalar theory, rather it is a mixed one having
vector fields and scalar fields as
basic structural terms. Moreover the generating function $\Psi$ is not a basic structural term, it is a derived quantity.\\
Since we want to establish a theory which has no other structural
terms than $\Psi, \tilde{\eta}$ and $\tilde{\vartheta}$ the
following heuristic idea is helpful: write down the equations of
motion and the Einstein equation in terms of the tetrad components
$\Lambda^\alpha_\beta$ for arbitrary coordinates, and in terms of
density $\eta$ and temperature $\vartheta$.

Then remove in $\Lambda^\alpha_\beta (x) = \frac{\partial
\phi^\alpha}{\partial \chi^\beta} (x, z) |_{z=x}$ the restriction
$x = z$, i.e. substitute
\begin{equation*}
\Lambda^\alpha_\beta (x) \quad \text{by} \quad \Pi^\alpha_\beta
(x, z): = \frac{\partial \Phi^\alpha}{\partial x^\beta} (x, z),
\end{equation*}

and generalize
$\eta (x)$ by $\tilde{\eta} (x, z)$ and $\vartheta (x)$ by $\tilde{\vartheta} (x, z)$.\\
The equations thus gained are taken for the remaining axioms of
the theory $\Phi_{sc}$. This program is carried through more
detailed in  the next chapter.

\section{Generalized Field Equations}

\subsection{The tetrad form of the field equations}

In this section the field equations, i.e. the equation of
continuity, the balance of energy and momentum and Einstein's
equation are formulated in terms of the components
$\Lambda^\alpha_\kappa, {\Lambda^{-1}}^\lambda_\beta$ of
$\Theta^\alpha$ and $e_\beta, \alpha, \beta = 1, \cdots, 4$ with
respect to an arbitrary coordinate system $\chi$ (cf.\  Remark
(5.14)). These equations are obtained from the usual formulation
in terms of the $\chi$-components $g_{\alpha \beta}$ and
$v_\alpha$ by inserting (cf.\ Formulae (5.1) and
(5.2))

\begin{equation} \label{10.1}
g_{\alpha \beta} = \Lambda^\kappa_\alpha \Lambda^\lambda_\beta
\eta_{\kappa \lambda} \quad \text{and} \quad v_\alpha = -
\Lambda^4_\alpha .
\end{equation}

In what follows, for the sake of convenience the abbreviations

\begin{equation} \label{10.1-2}
\Lambda: = ((\Lambda^\alpha_\beta)), \quad
\bar{\Lambda}^\alpha_\beta : = {\Lambda^{-1}} ^\alpha_\beta \quad
\text{and} \quad
 \bar\Lambda = ((\bar{\Lambda}^\kappa_\lambda)) = \Lambda^{-1}
\end{equation}

are used. Then the following proposition holds.\\

{\bf Proposition (10.1):} The equation of continuity,
$\mathrm{div} (\eta v) = 0$, reads:

\begin{equation} \label{prop10.1}
\bar \Lambda^\beta_4  \frac{\partial \eta}{\partial x^\beta} +
\eta (\bar{\Lambda}^\beta_4 \bar{\Lambda}^\alpha_\sigma -
\bar{\Lambda}^\beta_\sigma \bar{\Lambda}^\alpha_4)
\frac{\partial}{\partial x^\beta} \Lambda^\sigma_\alpha = 0
\end{equation}

The {\bf proof} is based on  the formula:

\begin{equation} \label{proof10.1}
\Gamma^\alpha_{\beta\gamma} = \left[ {\alpha \atop {\beta\gamma} }
\left. \right|   {\kappa \lambda \atop \sigma}     \right]
\frac{\partial}{\partial x^\kappa} \Lambda^\sigma_\lambda
\end{equation}

where

\begin{eqnarray*}
\begin{array}{c}
2 \left[ {\alpha \atop {\beta \gamma}}   \mid  { {\kappa \lambda}
\atop \sigma} \right]  =
\bar{\Lambda}^\alpha_\sigma  \left( \delta^\kappa_\beta \delta^\lambda_\gamma + \delta^\lambda_\beta \delta_\gamma^\kappa   \right) %\\ [0,5 cm]
  - \bar{\Lambda}^\alpha_\nu \bar{\Lambda}^\kappa_\varrho \eta^{\nu \varrho} \eta_{\sigma \mu}
\left( \Lambda^\mu_\gamma \delta^\lambda_\beta + \Lambda^\mu_\beta
\delta^\lambda_\gamma \right) \\ [0,5 cm]
  + \bar{\Lambda}^\alpha_\nu \bar{\Lambda}^\lambda_\varrho \eta^{\nu \varrho} \eta_{\sigma \mu}
\left(\Lambda^\mu_\gamma \delta^\kappa_\beta + \Lambda^\mu_\beta
\delta^\kappa_\gamma  \right).
\end{array}
\end{eqnarray*}

It can be derived by a straightforward but lengthy calculation.\\

Since the balance of energy and momentum depends strongly on the
constitutive equation

\begin{equation} \label{proof10.1-2}
T = {\mathcal{T}} (g, v, \eta, \vartheta)
\end{equation}

it cannot be written down explicitely for all the different
functionals $\mathcal{T}$. One case of major interest is that of a
Eulerian or ideal fluid. In this case $T$ is given by its
components:

 \begin{equation} \label{proof10.1-3}
T_{\alpha \beta}=  p \; g_{\alpha \beta} + h \; v_\alpha v_\beta
\end{equation}

where $p = \hat{p} (\eta,  \vartheta)$ and $h = \hat{h} (\eta, \vartheta)$.\\

{\bf Proposition (10.2):} The balance of energy and momentum,
$\mathrm{div} (T) = 0$, reads in terms of $\Lambda$ for a Eulerian
fluid:

\begin{equation} \label{prop10.2}
\begin{array}{cc}
{\bar{\Lambda}^\beta_j \frac{\partial p}{\partial x^\beta} + h
(\bar{\Lambda}^\kappa_4 \bar{\Lambda}^\beta_j -
\bar{\Lambda}^\beta_4 \bar{\Lambda}^\kappa_j)
\frac{\partial}{\partial x^\beta} \Lambda^4_\kappa = 0}, \quad j =
1, 2, 3  \\ [0,4 cm] {\bar{\Lambda}^\beta_4 \frac{\partial
p}{\partial x^\beta} - \bar{\Lambda}^\beta_4 \frac{\partial
h}{\partial x^\beta} - h (\bar{\Lambda}^\beta_4
\bar{\Lambda}^\kappa_\sigma - \bar{\Lambda}^\kappa_4
\bar{\Lambda}^\beta_\sigma) \frac{\partial}{\partial x^\beta}
\Lambda^\sigma_\kappa = 0}.
\end{array}
\end{equation}

Again, the {\bf proof} is based on (\ref{proof10.1}) together with some lengthy calculations.\\

Since the right-hand side of Einstein's equation (in its usual
form!) depends on the constitutive equation (\ref{proof10.1-2})
one can write
down only the left-hand side in an explicite way.\\
{\bf Proposition (10.3):} The Einstein equation $G + \Lambda_0 g =
\kappa_0 T$ reads:

\begin{equation} \label{prop10.3}
E^{\kappa \lambda \sigma}_{j k \mu} \left(\frac{\partial}{\partial
x^\lambda} \frac{\partial}{\partial x^\kappa} \Lambda^\mu_\sigma
\right) + D^{\kappa \varrho \lambda \sigma}_{j k \mu \nu} \left(
\frac{\partial}{\partial x^\kappa} \Lambda^\mu_\varrho \right)
\left(\frac{\partial}{\partial x^\lambda} \Lambda^\nu_\sigma
\right) + \Lambda_0 \eta_{\kappa \lambda} \Lambda^\kappa_j
\Lambda^\lambda_k = \kappa_0 T_{j k}
\end{equation}

where all indices run from 1 to 4. Here $\Lambda_0$ is an
(unspecified) cosmological constant and $\kappa_0$ is Einstein's
gravitational constant as usual. The coefficients $E$:: and $D$::
are explicitely given. They are
polynominals in $\Lambda$: and its inverse $\bar{\Lambda}$:.\\

The {\bf proof} is extremely lengthy, but also straightforward.\\

For later purposes it be noticed that the components of the Ricci
tensor have a similar form as (10.8). They read

\begin{equation} \label{proof10.3}
R_{j k} = S^{\kappa \lambda \sigma}_{j k \mu} \left(
\frac{\partial}{\partial x^\lambda} \frac{\partial}{\partial
x^\kappa} \Lambda^\mu_\sigma\right) + T^{\kappa \varrho \lambda
\sigma}_{j k \mu \nu} \left(\frac{\partial}{\partial x^\kappa}
\Lambda^\mu_\varrho\right) \left( \frac{\partial}{\partial
x^\lambda} \Lambda^\nu_\sigma \right).
\end{equation}

Hence

\begin{equation} \label{proof10.3-1}
E^{\kappa \lambda \sigma}_{j k \mu} = S^{\kappa \lambda \sigma}_{j
\varrho \mu} - \frac{1}{2} \eta_{\alpha \beta} \Lambda^\alpha_j
\Lambda^\beta_k \eta^{\iota \gamma} \Lambda^n_\iota
\Lambda^m_\gamma S^{\kappa \lambda \sigma}_{n m \mu}
\end{equation}

and similarly for $D$:: and $T$::.

\subsection{The generalizing procedure}

In this section the heuristic ideas presented at the end of
Section 9.3 are to be worked out in detail. For this purpose let
us again consider a chart $(V, \chi)$, and let the terms
$\Lambda^\alpha_\beta, \eta, \vartheta$ be functions of $x = \chi
(p), p \in V$. Moreover, let $\Phi (x, z): = \Psi (\chi^{-1} (x),
\chi^{-1} (z))$ for all $x, z \in  \chi [V]$ for which the
right-hand side is defined. Finally it is assumed that the
constitutive equation (\ref{proof10.1-2}) is given in the form

\begin{equation} \label{chap10.2}
T_{\alpha \beta} = {\mathcal{T}}'_{\alpha \beta} (\Lambda, \eta,
\vartheta).
\end{equation}

Especially for a Eulerian fluid it follows from
(\ref{proof10.1-3})  that

\begin{equation} \label{chap10.2-1}
T_{\alpha \beta} = p \Lambda^\kappa_\alpha \Lambda^\lambda_\beta
\eta_{\kappa \lambda} + h \Lambda^4_{\alpha} \Lambda^4_\beta.
\end{equation}

Then the generalized field equations are obtained from
(\ref{prop10.1}), (\ref{prop10.3}) and from $\mathrm{div} (T) =
0$, e.g. from (\ref{prop10.2}), by omitting the restriction $z =
x$ in $\Lambda (x) = \left. \frac{\partial}{\partial x} \Phi (x,
z) \right|_{z = x}$. More precisely, this means we have
to carry out in (\ref{prop10.1}), (\ref{prop10.3}) and (\ref{prop10.2}) (or more general in $\mathrm{div} (T) = 0)$ the following\\

{\bf Substitution (10.4):}\\
\begin{enumerate}
\item[1.]
\begin{equation}  \label{sub10.4}
\Lambda (x) = \left. \frac{\partial}{\partial x} \Phi (x, z)
\right|_{z = x} \longrightarrow \Pi (x, z): =
\frac{\partial}{\partial x} \Phi (x, z)
\end{equation}

\begin{equation} \label{sub10.4-1}
\bar{\Lambda} (x): = \Lambda^{-1} (x) \longrightarrow \bar{\Pi}
(x, z): =  \Pi^{-1} (x, z).
\end{equation}

\item[2.]

\begin{equation} \label{sub10.4-2}
\eta (x) \longrightarrow \tilde{\eta} (x, z)
\quad\mathrm{and}\quad \vartheta (x) \longrightarrow
\tilde{\vartheta} (x, z)
\end{equation}
 where

\begin{equation} \label{sub10.4-3}
\tilde{\eta} (x, x) = \eta (x) \quad \mathrm{and}\quad
\tilde{\vartheta} (x, x) = \vartheta (x).
\end{equation}

\item[3.] For the derivatives of $\Lambda$ it is natural to set

\begin{equation} \label{sub10.4-4}
\frac{\partial}{\partial x^\varrho} \Lambda^\kappa_\alpha
\longrightarrow \frac{\partial}{\partial x^\varrho}
\Pi^\kappa_\alpha + \frac {\partial}{\partial z^\varrho}
\Pi^\kappa_\alpha = \frac{\partial}{\partial x^\varrho}
\frac{\partial}{\partial x^\alpha} \Phi^\kappa +
\frac{\partial}{\partial z^\varrho} \frac{\partial}{\partial
x^\alpha} \Phi^\kappa,
 \end{equation}

\begin{equation} \label{sub10.4-5}
\begin{array}[t]{lll}
{\dfrac{\partial}{\partial x^\sigma} \dfrac{\partial}{\partial
x^\varrho} \Lambda^\kappa_\alpha} &\longrightarrow &
\dfrac{\partial}{\partial x^\sigma} \dfrac{\partial}{\partial
x^\varrho} \Pi^\kappa_\alpha  + \dfrac {\partial}{\partial
z^\sigma} \dfrac {\partial}{\partial x^\varrho} \Pi^\kappa_\alpha
\\  [0.4cm]&+& {\dfrac{\partial}{\partial x^\sigma}
\dfrac{\partial}{\partial z^\varrho} \Pi^\kappa_\alpha +
\dfrac{\partial}{\partial z^\sigma} \dfrac{\partial}{\partial
z^\varrho}  \Pi^\kappa_\alpha} \\ [0.4cm] &=&
{\dfrac{\partial}{\partial x^\sigma} \dfrac{\partial}{\partial
x^\varrho} \dfrac{\partial}{\partial x^\alpha} \Phi^\kappa +
\cdots + \dfrac{\partial}{\partial z^\sigma}
\dfrac{\partial}{\partial z^\varrho} \dfrac{\partial}{\partial
x^\alpha} \Phi^\kappa}
\end{array}
\end{equation}

\item[4.] Likewise the derivatives of $\eta$ and $\vartheta$ are replaced by

\begin{equation} \label{sub10.4-6}
\frac{\partial}{\partial x^\alpha} \eta \longrightarrow
\frac{\partial}{\partial x^\alpha} \tilde{\eta} + \frac
{\partial}{\partial z^\alpha} \tilde\eta, \quad
 \frac{\partial}{\partial x^\alpha} {\vartheta} \longrightarrow
\frac{\partial}{\partial x^\alpha} \tilde{\vartheta} +
 \frac{\partial}{\partial z^\alpha} \tilde{\vartheta}
 \end{equation}
\end{enumerate}

{\bf Remark (10.5):}  By definition we have $\Pi (x, x) = \Lambda
(x), \tilde{\eta} (x, x) = \eta (x)$ and $\tilde{\vartheta} (x, x)
= \vartheta (x)$. The same holds for (\ref{sub10.4-4}),
(\ref{sub10.4-5}) and (\ref{sub10.4-6}). This means, if $z = x$ on
the right-hand side the arrow $\rightarrow$
can be replaced by =.\\

\subsection{Results}
{\bf 10.3.1:} Applying the rules of Substitution (10.4) to Equation (\ref{prop10.1}) we arrive at the following\\
{\bf Proposition (10.6):} The generalization of the equation of
continuity yields

\begin{equation*}
{\mathcal{C}} (\Phi, \tilde{\eta}) = 0
\end{equation*}
where

\begin{equation} \label{prop10.6}
{\mathcal{C}} (\Phi, \tilde{\eta}) = \bar{\Pi}^\beta_4
\left(\frac{\partial \tilde{\eta}}{\partial x^\beta} +
\frac{\partial \tilde{\eta}}{\partial z^\beta} \right) +
\tilde{\eta} \left(\bar{\Pi}^\beta_4 \bar{\Pi}^\kappa_\lambda -
\bar{\Pi}^\kappa_4 \bar{\Pi}^\beta_\lambda \right)
\frac{\partial}{\partial z^\beta} \frac{\partial}{\partial
x^\kappa} \Phi^\lambda .
\end{equation}

The {\bf proof} is trivial.\\

Since the generalizing procedure cannot be carried through for
arbitrary constitutive equations (\ref{chap10.2}) we confine again
to
the case of a Eulerian fluid. \\
{\bf Proposition (10.7):} The generalization of the balance of
energy and momentum yields for a Eulerian fluid
\begin{equation*}
{\mathcal{E}}_\varrho (\Phi, \tilde{\eta}, \tilde{\vartheta}) = 0,
\quad \varrho = 1,  \cdots, 4
\end{equation*}
 where

\begin{equation} \label{prop10.7}
{\mathcal{E}}_n (\Phi, \tilde{\eta},  \tilde{\vartheta}) =
\bar{\Pi}^\beta_n \left(\frac{\partial \tilde{p}}{\partial
x^\beta} + \frac{\partial \tilde{p}}{\partial z^\beta} \right) -
\tilde{h} \left(\bar{\Pi}^\beta_4 \bar{\Pi}^\kappa_n -
\bar{\Pi}^\kappa_4 \bar{\Pi}^\beta_n \right)
\frac{\partial}{\partial z^\beta}  \frac{\partial}{\partial
x^\kappa} \Phi^4
\end{equation}

for $n = 1, 2, 3$ and

\begin{eqnarray} \label{prop10.7-1}
{\mathcal{E}}_4 (\Phi, \tilde{\eta},  \tilde{\vartheta}) & =
\bar{\Pi}^\beta_4 \left(\dfrac{\partial \tilde{h}}{\partial
x^\beta} - \dfrac{\partial \tilde{p}}{\partial x^\beta} +
\dfrac{\partial \tilde{h}}{\partial z^\beta} - \dfrac{\partial
\tilde{p}}{\partial z^\beta}  \right)
\\
&+ \tilde{h} \left(\bar{\Pi}^\beta_4 \bar{\Pi}^\kappa_\lambda -
\bar{\Pi}^\kappa_4 \bar{\Pi}^\beta_\lambda \right)
\dfrac{\partial}{\partial z^\beta}  \dfrac{\partial}{\partial
x^\kappa} \Phi^\lambda; \nonumber
\end{eqnarray}

moreover, using Equation (\ref{proof10.1-3}) we have

\begin{equation} \label{prop10.7-2}
\tilde{p} = \hat{p} \, (\tilde{\eta}, \tilde{\vartheta}) \;
\mathrm{and} \; \tilde{h} = \hat{h} (\tilde{\eta},
\tilde{\vartheta}).
\end{equation}

The {\bf proof} is again simple.\\

Finally the replacement rules are applied to Equation (\ref{prop10.3}).\\

{\bf Proposition (10.8):} If the replacement rules applied to the
constitutive equation (\ref{chap10.2}) make sense, the
generalization of the Einstein equation yields

\begin{equation} \label{prop10.8}
{\mathcal{G}}_{jk} + \Lambda_0 \Pi^\lambda_j \Pi^\kappa_k
\eta_{\lambda \kappa} = \kappa_0 \tilde{T}_{j k}
\end{equation}

where

 \begin{equation} \label{prop10.8-1}
\tilde{T}_{j k} = {\mathcal{T}}'_{jk} (\Pi, \tilde{\eta},
\tilde{\vartheta})
\end{equation}

and

 \begin{eqnarray} \label{prop10.8-2}
{\mathcal{G}}_{j k} (\Phi) & = {\stackrel{1}{M}}^{\lambda \kappa
\sigma}_{j k \mu} \dfrac{\partial^3 \Phi ^\mu}{\partial z^\lambda
\partial z^\kappa \partial x^\sigma} + {\stackrel{2}{M}}^{\lambda
\kappa \sigma}_{j k \mu}
\dfrac{\partial^3 \Phi^\mu}{\partial z^\lambda \partial x^\kappa \partial x^\sigma} \\[0.5 cm]
& + {\stackrel{1}{K}}^{\lambda \kappa \varrho \sigma}_{j k \mu
\nu} \left(\dfrac{\partial^2 \Phi^\mu}{\partial z^\lambda \partial
x^\kappa} \right) \left(\dfrac{\partial^2 \Phi^\nu}{\partial
z^\varrho \partial x^\sigma} \right) \nonumber \\[0.5 cm]
& + {\stackrel{2}{K}}^{\lambda \kappa \varrho \sigma}_{j k \mu
\nu} \left(\dfrac{\partial^2 \Phi^\mu}{\partial z^\lambda \partial
x^\kappa} \right) \left(\dfrac{\partial^2 \Phi^\nu}{\partial
x^\varrho \partial x^\sigma} \right) \nonumber
\end{eqnarray}

(All indices run from 1 to 4.) Again the {\bf proof} is simple but very lengthy.\\

The properties of the substituted terms mentioned in Remark (10.5) have some consequences which are important later on.\\

{\bf Remark (10.9):} It follows from the construction of the terms
${\mathcal{C}}, {\mathcal{E}}_\varrho$ and ${\mathcal{G}}_{j k}$,
$\varrho, j, k = 1, \cdots, 4$ that

\begin{equation} \label{rem10.9}
{\mathcal{C}} (\Phi, \tilde{\eta}) (x, x) = \mathrm{div} (\eta v)
(x)
\end{equation}

where $\mathrm{div} (\eta v) (x)$ is the left-hand side of (10.3),
and that

\begin{equation} \label{rem10.9-1}
{\mathcal{G}}_{jk} (\Phi) (x, x) = G_{j k} (x) = R_{j k} (x) -
\frac{1}{2} g_{j k} (x) \bar{R} (x)
\end{equation}

where $R_{jk}$ is given by (10.9) and $G_{jk}$ by the first and the second term of the left-hand side of (10.8).\\
 Moreover, ${\mathcal{E}}_\varrho (\Phi, \tilde{\eta}, \tilde{\vartheta})
(x, x)$ is equal to the left-hand side of (\ref{prop10.2}) \\

{\bf 10.3.2:} In this section the results of the Propositions
(10.6) to (10.8) are generalized
further. In order to do this let us introduce the following\\

{\bf Notation (10.10):} For a given chart $(V, \chi)$ let
${\mathcal{V}} \subset \chi [V] \times \chi [V] $ be the domain of
$\Phi = \Psi (\chi^{-1}, \chi^{-1})$. Then a real function
$\mathcal{O}$ defined on $\mathcal{V}$ such that ${\mathcal{O}}
(x, x) = 0$,
is called {\it quasi-zero}.\\
Now the following proposition holds:

{\bf Proposition (10.11):} Let the functions $\Phi, \tilde{\eta},
\tilde{\vartheta}$ be defined on $\mathcal{V}$, and let
${\mathcal{C}}, {\mathcal{E}}_\varrho, {\mathcal{G}}_{jk}$ be
given by the Formulae (\ref{prop10.6}), (\ref{prop10.7}),
(\ref{prop10.7-1}) and (\ref{prop10.8-2}). If $\Phi, \tilde{\eta},
\tilde{\vartheta}$ satisfy any subset of the equations

\begin{equation} \label{prop10.11}
{\mathcal{C}} (\Phi, \tilde{\eta}) = {\mathcal{O}},
\end{equation}

\begin{equation} \label{prop10.11-1}
{\mathcal{E}}_\varrho (\Phi, \tilde{\eta}, \tilde{\vartheta}) =
{\mathcal{O}}_\varrho
\end{equation}

\begin{equation} \label{prop10.11-2}
{\mathcal{G}}_{jk} (\Phi) + \Lambda_o \Pi^\lambda_j \Pi^\kappa_k
\eta_{\lambda k} = \kappa_0 {\mathcal{T}}'_{jk} (\Pi,
\tilde{\eta}, \tilde{\vartheta}) + {\mathcal{O}}_{jk},
\end{equation}

where ${\mathcal{O}}, {\mathcal{O}}_\varrho$ and
${\mathcal{O}}_{jk}, \varrho, j, k = 1, \cdots, 4$ are
quasi-zeros, then the terms $\Lambda, \eta, \vartheta$ defined by
$\Lambda (x) = \Pi (x, x), \; \eta (x) = \tilde{\eta} (x, x), \;
\vartheta (x) = \tilde{\vartheta} (x, x)$ satisfy the
corresponding subset
of the Equations (\ref{prop10.1}), (\ref{prop10.2}) and (\ref{prop10.3}).\\
The {\bf proof} results immediately from Remark (10.9).\\

{\bf Notation (10.12):} The equations (\ref{prop10.11}) to
(\ref{prop10.11-2}) are called the generalized equation of
continuity, the generalized balance of
energy and momentum and the generalized Einstein equation.\\
{\bf Remark (10.13):} If one has a solution $\Phi, \tilde{\eta},
\tilde{\vartheta}$ of the generalized Einstein equation such that
$\tilde{\eta}$
is a quasi-zero, then $\Lambda$ is a vacuum solution of Einstein's equation.\\

To a certain extent also the converse of Proposition (10.11)
holds:

{\bf Proposition (10.14):} Let be given a solution $\Lambda, \eta,
\vartheta$ of the Equations (10.3) and (10.8), i.e. of the
equation of continuity and of Einstein's equation in tetrad form.
Moreover, let $\Phi$ be given by Definition (8.12) of Part II and
define $\tilde{\eta}, \tilde{\vartheta}$ by

\begin{displaymath} \tilde{\eta}(x, z) = \eta(z^{1}, z^{2}, z^{3}, x^{4}) ,\;  \tilde{\vartheta}(x, z) = \vartheta(z^{1}, z^{2}, z^{3}, x^{4}) \end{displaymath}

Finally, let the energy-momentum tensor $T$ be defined by a
functional $\mathcal{T}'$ such that for the generalization
$\tilde{T}$ of $T$ the following relations hold for $j, k = 1,
...., 4:$

\begin{displaymath}\tilde T_{jk}(x, z) = {\mathcal{T}}'_{jk}(\Pi, \tilde \eta , \tilde \vartheta) (x, z) = {\mathcal{T}}'_{jk}(\Lambda, \eta, \vartheta)(\bar{z},x^{4}) + {\mathcal{O}}_{jk}\end{displaymath}

with ${\mathcal O}_{jk}$ being quasi-zeros and $\bar{z} = (z^{1},
z^{2}, z^{3})$.
Then the triple $ \Phi, \tilde \eta , \tilde \vartheta $ is a (local) solution of the generalized equations (10.29) and (10.31).\\
The \textbf{Proof} is again very lengthy but straightforward.

\section{The Theory $\Phi_{sc}$}
\subsection{Geometry and Kinematics}

In this chapter the theory $\Phi_{sc}$ is formulated in the sense
of Section 1.2 (i) of Part I, i.e. the terminology of Section 2.2
is used. Since  the inductive procedure for setting up $\Phi_{sc}$
is described extensively in the Chapters 9 and 10 it suffices to
write down the elements of $\Phi_{sc}$ without any
further comment.\\

The {\it base sets} of $\Phi_{sc}$ are $M$ and $\mathbb{R}$, and
$\Psi, \tilde{\eta}, \tilde{\vartheta}$ are its \textit{
structural terms}.

The physical interpretation of these terms is this:\\
 $M$ is the set of signs for events;\\
 $\Psi$ determines an atlas of pre-radar charts;\\
$\tilde{\eta}$ is a generalized density which determines the density $\eta$ by $\tilde{\eta} (p, p) = \eta (p)$; \\
$\tilde{\vartheta}$ is a generalized temperature which determines
the (empirical)
temperature $\vartheta$ by $\tilde{\vartheta} (p, p) = \vartheta (p)$.\\
At first the axioms for geometry and kinematics are formulated:\\

{\bf GK$_{sc}$1:} $\Psi$ is a function, $\Psi: {\mathcal{M}}
\longrightarrow \mathbb{R}^4$ where ${\mathcal{M}}: = \cup_{q \in
M} V_q \times \{q\}$
with $V_q \subset M, \quad V_q \not= \emptyset$ and $q \in V_q$.\\

{\bf GK$_{sc}$2:}  The term ${\mathcal{A}}: = \{(V_q, \Psi (\cdot,
q)) : q \in M\}$ is a $C^k$-atlas, $k \ge 3$ on $M$ so that $(M,
\mathcal{A})$
is a connected Hausdorff manifold. \\

{\bf GK$_{sc}$3:} $\Psi$ is of class $C^k, k \ge 3$ \\

{\bf GK$_{sc}$4:}  For each $q \in  M$ there is a maximal open
interval $J_q$ such that $(0, 0, 0,  \tau)  \in \mathrm{ran}  \Psi
(\cdot, q)$
for each $\tau \in J_q$. \\

It is useful to introduce the following\\
{\bf Notation (11.1):} 1. For each $q \in M$ the function
$\gamma_q : J_q \longrightarrow M$ is defined by $\gamma_q (t) =
\Psi (\cdot, q)^{-1}
 (0, 0, 0, t)$. Moreover, we write $W_q: = \mathrm{ran} \gamma_q$.\\
2. ${\mathcal{D}}$ is the differential structure which contains
all charts $(V, \chi)$ which are $C^k$-compatible, $k \ge 3$, with
$\mathcal{A}$.\\

{\bf GK$_{sc}$5:}  For each $q \in M$ there is a $t_q \in J_q$ such that $\gamma_q (t_q) = q$.\\

{\bf GK$_{sc}$6:} For all $q' \in W_q$ it holds that $\Psi (\cdot, q') = \Psi (\cdot, q)$.\\

\subsection{Field equations}
In the next step the axioms of the motion of matter are
formulated. For this purpose a notation is used which depends on
coordinates. Moreover, for any function $F$ depending on $p, q \in
M$ and its coordinate form the same symbol is used, i.e. we write
$F (p, q) = F (x, z)$ for $ x = \chi (p)$ and $z = \chi (q)$. \\

{\bf EM$_{sc}$1:}{ The terms $\tilde{\eta}$ and
$\tilde{\vartheta}$ are functions, $\tilde{\eta}: {\mathcal{M}}
\rightarrow \mathbb{R}$, $\tilde{\vartheta}: {\mathcal{M}}
\rightarrow \mathbb{R}$
which are of class $C^r, r \ge 2$ and where ${\mathcal{M}}$ is the set introduced in GK$_{sc}1$.} \\

{\bf EM$_{sc}$2:} { Let $(V, \chi) \in {\mathcal{D}}$ be any chart
and let $(x, y) \in  {\mathcal{V}}:= (\chi \times \chi)
[{\mathcal{M}} \cap (V \times V)]$. Then the $\chi$-components of
the generalized energy-momentum tensor $\tilde{T}_{jk}$ are given
by

\begin{equation*}
\tilde{T}_{jk} (x, z) = {\mathcal{T}}'_{jk} (\Pi, \tilde{\eta},
\tilde{\vartheta}) (x, z)
\end{equation*}

where ${\mathcal{T}}'_{jk}$ is the same functional as in Formula
(\ref{chap10.2}) for which Substitution (10.4) makes sense.\\
[0,1cm]

{\bf EM$_{sc}$3:} For each chart $(V, \chi) \in {\mathcal{D}}$
there is a quasi-zero $\mathcal{O}$ such that the equation
${\mathcal{C}} (\Phi, \tilde{\eta}) = \mathcal{O}$ holds.\\

{\bf EM$_{sc}$4:} For each chart $(V, \chi) \in {\mathcal{D}}$
there are quasi-zeros ${\mathcal{O}}^\alpha, \alpha = 1, \cdots,
4$ so that the generalization of the balance of energy and
momentum holds for $\tilde{T}^{\alpha \beta}$ and with
${\mathcal{O}}^\alpha$ at the right-hand side. Especially, for
Euler fluids the resulting equation reads ${\mathcal{E}}_\varrho
(\Phi, \tilde{\eta}, \tilde{\vartheta}) = {\mathcal{O}}_\varrho,
\varrho = 1, \cdots, 4$. where ${\mathcal{O}}_\varrho$ are
quasi-zeros and where ${\mathcal{E}}_\varrho$ is defined in
Proposition (10.7).\\ [0,1cm]

Likewise, the generalized Einstein equation is introduced by an axiom:\\

{\bf EE$_{sc}$:} For each chart $(V, \chi) \in {\mathcal{D}}$
there are quasi-zeros ${\mathcal{O}}_{jk}, j, k = 1, \cdots, 4$ so
that the equations

\begin{equation*}
 {\mathcal{G}}_{jk} (\Phi) + \Lambda_0 \Pi^\lambda_j \Pi^\kappa_k \eta_{\lambda \kappa} =
\kappa_0 {\mathcal{T}}'_{jk} (\Pi, \tilde{\eta},
\tilde{\vartheta}) + {\mathcal{O}}_{jk}
\end{equation*}

hold where $\Lambda_0$ is an unspecified cosmological constant and where $\kappa_0$ is Einstein's gravitational constant.\\

\subsection{Additional Conditions}
Finally, in order to complete the axioms of $\Phi_{sc}$ the
additional conditions (AC) have to be formulated.
As in Section 4.3 we only illustrate the subject by three examples:\\
(i)  Initial conditions;\\
(ii)  Boundary conditions;\\
(iii) Symmetry conditions.

In Section 12.2 we come back to the fomulation of symmetry conditions for the function $\Psi$.\\

\subsection{Some consequences of $\Phi_{sc}$}
{\bf 11.4.1:} If one defines the tetrads $\Theta^\alpha$ and
$e_\beta, \alpha, \beta = 1, \cdots, 4$ by

\begin{equation*}
\Theta^\alpha (p) = d \Psi^\alpha (\cdot, p) |_p \quad \text{and}
\quad e_\beta (p) = \partial_{{\Psi^\beta}(\cdot, p)} |_p
\end{equation*}

and the metric $g$ and the velocity $v$ by

\begin{equation*}
g = \eta_{\alpha \beta} \Theta^\alpha \otimes \Theta^\beta   \quad
\text{and} \quad v = e_4,
\end{equation*}

then $\Psi$ is a full generating function in the sense of
Definition (7.2), i.e.\ $\Psi$ satisfies the Conditions P1 to P5
of Section 7.1.2. This follows directly from the Axioms GK$_{sc}$1
to 6 and from Proposition (9.1). If one is interested only in a
partial generating function $\Psi$ satisfying the Conditions P1 to
P3 then $\Phi_{sc}$ can be weakened by omitting the Axioms
GK$_{sc}$ 4 to 6.

{\bf 11.4.2:} Let $\gamma_q$ be defined by Notation (11.1). Then
$\gamma_q$ is bijective and of class $C^k, k \ge 3$. This follows
directly from $\Psi (\cdot, q) \circ \gamma_{q} (t) = (0, 0, 0, t)$. \\

{\bf 11.4.3:} From Axiom GK$_{sc}$5 and from Notation (11.1) it
follows that $q \in W_q$ for each $q \in M$, and therefore
$\cup_{q \in M} W_q = M$. Moreover, from Axiom GK$_{sc}$6 we
conclude that $\gamma_{q'} = \gamma_q$ for $q' = W_{q}$. Hence
$J_{q'} = J_{q}$
and $W_{q'} = W_{q}$ for $q' \in W_{q}.$\\

{\bf 11.4.4:} If $q' \notin W_{q}$ then $W_{q} \cap W_{q'} =
\emptyset$. For assume that $\bar{q} \in W_q \cap W_{q'}$, then
$\bar{q} \in W_{q}$
and $\bar{q} \in W_{q'}$, so that $W_{q} = W_{\bar{q}} = W_{q'}$.\\

{\bf 11.4.5:} For each $q \in M$ let us select exactly one element
$\hat{q}$ from $W_q$ so that $\hat{q}$ is also selected from
$W_{q'}$ for $q' \in W_q$, and let $N \subset M$ be the set of all
these selected $\hat{q}$. Now, let $P$ be any set of the same
cardinality as $N$. Then, identifying a particle in $\Phi_{sc}$
with a worldline $W_q$ the set $P$ is a set of indices for
particles as is used in the theories $\Phi_R$ and $\Phi^\ast_R$
(cf.\ Notation (2.1) and (6.1)). Let $A \in P$ denote a particle
and let $A \leftrightarrow \hat{q}$. Then by $\psi_A : = \Psi
(\cdot, \hat{q}), \gamma_A := \gamma_{\hat{q}}$ and $W_A: =
W_{\hat{q}}$ the notation used in $\Phi_R$ and $\Phi^\ast_R$ is
regained. Also the function $F$ (cf.\ Remark (3.6)) is
defined in $\Phi_{sc}$ by $F(q) = A$ for all $q \in W_A$. \\

\subsection{Remarks concerning models of $\Phi_{sc}$}
By Notation (4.1) of Part I the concept of a model was explicetly
introduced for the theory $\Phi_R$. It can be transfered quite
easily to each theory $\tilde{\Phi}$ which is formulated according
to the scheme (i) in the Sections
1.2 and 2.2 as follows:\\

{\bf Notation (11.2):} If one replaces the base sets and the
structural terms of $\tilde{\Phi}$ by explicit terms of
mathematical analysis (or of the theory of sets) such that these
terms satisfy the axioms of $\tilde{\Phi}$ within mathematical
analysis (or the theory of sets),
then we say that these termes define an analytical (or a set theoretical) model of $\tilde \Phi$.\\
In the usual formulations of GR the additional conditions (AC) are
chosen such that the models are unique (or unique up to some
diffeomorphisms). In case of the theory $\Phi_{sc}$ the situation
is different. Uniqueness is not needed for the models. Rather the
generalized density $\tilde \eta$ and the generalised temperature
$ \tilde \vartheta$ only have to be unique up to quasi-zeros. Then
different $\tilde \eta$ and $\tilde \vartheta$ define the same
physically interpretable fields $\eta$ and $\vartheta$ by $\tilde
\eta (p, p) = \eta (p)$ and $\tilde \vartheta (p, p) = \vartheta
(p)$. Likewise, the function $\Psi$ need not be unique. Any two
model functions $\Psi$ and $\Psi'$ describe the same physical
situation if they generate
the same differential structure $\mathcal{D}$, the same metric $g$ and the same velocity $v$. This suggests the following\\

{\bf Definition (11.3):} Any two arrays of terms $M, \Psi, \tilde
\eta, \tilde \vartheta$ and $M, \Psi', \tilde{\eta}',
\tilde{\vartheta}'$ forming models of the theory $\Phi_{sc}$ are
called {\it physically equivalent} if $\Psi$ and $\Psi'$ generate
the same ${\mathcal{D}}, g$ and $v$
and if $\tilde\eta, \tilde\vartheta$ and $\tilde{\eta} ', \tilde{\vartheta}'$ differ only by a quasi-zero. \\

Clearly, physical equivalence is an equivalence relation within
the models of $\Phi_{sc}$. Consequently, the axioms AC should be
such that they determine uniquely a class of physically equivalent
models of $\Phi_{sc}$. But up to now, the mathematical question is
still open how to formulate a well-posed initial value problem for
the generalized Einstein equation and the generalized
equation of continuity so that an equivalence class is uniquely determined.\\

{\bf Remark (11.4):} 1. If $\Psi$ and $\Psi'$ belong to two
physically equivalent models they both satisfy the Conditions P1
to P5 of Section 7.1.2 in Part II and are related by Formula
(7.8) where the Lorentz matrix $L$ and the function $R$
obey the relations
(7.12), (7.14), (7.15) and $d_p R(p, q)|_{q=p} = 0$.\\
This follows directly from Proposition (9.1) and the Propositions
(7.5), (7.11) and (7.12) together with Corollary (7.10).
Especially from Formula (7.14) one concludes that $R$
is a quasi-zero because there is a $t_q \in J_q$ such that
$\gamma_q (t_q) = q$ for each $q \in M$.\\
2. Since one is only interested in a class of physically
equivalent models the theory $\Phi_{sc}$ is a gauge theory.

\section{Further Properties of Generating Functions}

\subsection{Orientation and time orientation}

In this chapter a connected Hausdorff manifold $(M,
{\mathcal{A}}^+)$ is considered where ${\mathcal{A}}^+$  is of
class $C^k, k \ge 3$. Moreover, it is assumed that a partial
generating function $\Psi$ in the sense of Definition (7.2) is
defined on $M$. This implies that  $\Psi$ satisfies Condition P1
of Section 7.1.2, i.e.\ that the atlas ${\mathcal{A}}$ generated
by $\Psi$ is $C^k$-compatible with ${\mathcal{A}}^+$. In other
words, $\Psi$ generates a differential structure ${\mathcal{D}}$
which contains ${\mathcal{A}}^+$. These assumptions are satisfied
by the theories $\Phi_R$ and $\Phi_R^\ast$ treated in Part I and
the theory $\Phi_{sc}$ introduced in Chapter 11. Later on some
additional assumptions are needed. The above assumptions allow to
introduce the tetrads $\Theta^\alpha, e_\beta, \alpha, \beta = 1,
\cdots, 4$ in the usual way by

\begin{equation} \label{12}
\Theta^\alpha {(p)} = d \Psi^\alpha (\cdot, p) |_p \quad
\text{and} \quad e_\beta (p) = \partial_{\Psi^\beta (\cdot, p)}
|_p
 \end{equation}

as well as the fields $g$ and $v$ by (9.3). Then the following simple result holds: \\
{\bf Proposition (12.1):} The 4-form $\omega$ defined by

\begin{equation} \label{12-1}
\omega = \Theta^1 \wedge \Theta^2 \wedge \Theta^3 \wedge \Theta^4
\end{equation}

determines an orientation on $M$. \\

For the {\bf proof} one has to show that $\omega$ nowhere
vanishes. This follows directly from $\omega (e_1, \cdots, e_4)
=1$
throughout $M$.\\

For the next step we need the assumption that a Lorentz metric $g$
and a velocity $v$ is defined on $M$ and that
$g (v, v) = -1$.\\
Then the  manifold $(M, {\mathcal{A}}^+, g)$ is time orientable (cf.\ e.g.\ \cite{sachs77} p.\ 26).\\
In a further step it is assumed that $\Psi$ in addition satisfies
the conditions P2 and P3. This means that the differential
structure ${\mathcal{D}}$ defined by ${\mathcal{A}}$ is generated
by $\Psi$ and that $\Psi$ also generates $g$ and $v$. In this
case a time orientation is given by \\

{\bf Definition (12.2):} Let $u \in T_p M$ be timelike or
lightlike. Then $u$ is called future-pointing if
$\Theta^4 {(u)} = -g (v, u) > 0$ and past-pointing if $\Theta^4 (u) = - g(v,u) <0$.\\

{\bf Remark (12.3):} Independently of the fact that $(M,
{\mathcal{A}}^+, g)$ is time orientable if a velocity field $v$
exists on $M$ it can be seen that Definition (12.2) makes sense.
For, it can be shown that a future-pointing vector can not be
transferred into a past-pointing only by parallel transport, and vice versa.\\

\subsection{Isometries}

The aim of this section is defining the concept of isometry solely
in terms of a generating function $\Psi$, i.e.\ without using
explicitly a metric $g$. For this purpose we use again the
notation and the suppositions introduced at the
beginning of Section 12.1.\\

Moreover, troughout this section it is assumed that a bijective
function  $ f : M \rightarrow M$ is given. Then let $\Psi' := \Psi
(f,f)$. Finally, let the term ${\mathcal{A}}'$ be defined by

\begin{equation} \label{iso}
 {\mathcal{A}}' = \left\{(V'_{q'}, \Psi' (\cdot, q')) : q' \in M, V'_{q'} = f^{-1} [V_q], V_q = \mathrm{dom} \Psi (\cdot, q), q = f(q') \right\}.
\end{equation}

Then we obtain the following result:\\
{\bf Remark (12.4):} If $f$ is a homeomorphism then
${\mathcal{A}}'$ is a $C^k$-atlas, $k \ge 3$ on $M$. For,
$V'_{q'}$ is open because $V_q$ is open and $f$ is continuous.
Moreover, $\Psi' (\cdot, q') = \Psi (\cdot, f (q')) \circ f$ is a
homeomorphism. Hence $(V'_{q'}, \Psi' ( \cdot, q'))$ is a chart.
Each two charts are $C^k$-compatible because

\begin{equation} \label{iso-1}
\Psi' (\cdot, q'_1) \circ \Psi^{'-1} (\cdot, q'_2) = \Psi (\cdot,
f (q'_1)) \circ \Psi^{-1} (\cdot, f (q'_2)).
\end{equation}

In the next step it is assumed that $f$ is differentiable.\\

{\bf Proposition (12.5):} The function $f$ is a
$C^k$-diffeomorphism, $k \ge 3$ exactly if ${\mathcal{A}}$ and
${\mathcal{A}}'$ are
$C^k$-compatible atlases, $k \ge 3$.\\

{\bf Proof:} 1. If $f$ is diffeomorphic then ${\mathcal{A}}'$ is
an atlas. Let $\Psi' (\cdot, q'), \Psi (\cdot, q)$ be arbitrary
coordinate functions from ${\mathcal{A}}'$ resp. from
${\mathcal{A}}$. Then the function

\begin{equation} \label{iso-2}
\Psi' (\cdot, q') \circ \Psi^{-1} (\cdot, q) = \Psi (\cdot, f
(q')) \circ f \circ \Psi^{-1} (\cdot, q)
\end{equation}

is of class $C^k, k\ge 3$, and likewise its inverse, so that ${\mathcal{A}}$ and ${\mathcal{A}}'$ are $C^k$-compatible.\\
2. If  ${\mathcal{A}}$ and ${\mathcal{A}}'$ are $C^k$-compatible
atlases the left-hand side of (\ref{iso-2}) and its inverse are of
class $C^k$, so that
$f$ is a $C^k$-diffeomorphism.\\

In a last step the main result of this section is formulated. For
this purpose it is assumed that a metric $g$ is defined on the
manifold $(M, {\mathcal{A}}^+)$.  Then a diffeomorphism $f$ is
called an isometry if $g (p') = f^*_{p'} g (p)$ where $p = f(p')$
and
$f^*_{p'}$ is the pull back of $f$ at $p'$.\\

{\bf Proposition (12.6):} Let $\Psi$ satisfy the Conditions P1 and
P2 of Section 7.1.2, i.e.\ $\Psi$ generates the differential
structure ${\mathcal{D}}$ of class $C^k, k \ge 3$ containing
${\mathcal{A}}^+$, and the metric $g$. Then the (bijective)
function $f$ is a $C^k$-isometry, $k \ge 3$ exactly if $\Psi'$
satisfies also P1 and P2, i.e.\ if $\Psi'$ generates
${\mathcal{D}}$
and $g$.\\

{\bf Proof:} 1. First of all some auxiliary formulae are proved.
Let $f$ be a  diffeomorphism, and let $p = f (p')$ and $q =
f(q')$. Then

\begin{equation} \label{iso3}
d \Psi' (\cdot, q') |_{p'} = d \Psi (f, q) \mid_{p'} = f^\ast_{p'}
d \Psi (\cdot, q) |_p.
\end{equation}

This equation is seen to be true by the following short
calculation. Let $w' \in T_{p'} M$, then

\begin{equation} \label{iso4}
\begin{array}{ll}
d_{p'} \Psi^{'\alpha} (\cdot, q') (w') = w' (\Psi^{' \alpha}
(\cdot, q')) \\ [0.4cm] = w' (\Psi^\alpha (\cdot, q) \circ f) =
f_{*p'} w' (\Psi^\alpha (\cdot, q))  \\ [0.4cm] = d_p \Psi^\alpha
(\cdot, q) (f_{*p'} w') = f^*_{p'} d_p \Psi^\alpha (\cdot, q)
(w').
\end{array}
\end{equation}

Now let $g'$ be the metric which is generated by $\Psi'$:

\begin{equation} \label{iso5}
g' (p') := \eta_{\alpha \beta} d \Psi^{'\alpha} (\cdot, p') |_{p'}
\otimes d \Psi^{'\beta} (\cdot, p') |_{p'}.
\end{equation}

Then from (\ref{iso3}) it follows that

\begin{equation} \label{iso6}
g' (p') = f^\ast_{p'} g (p).
\end{equation}

2. If $f$ is a $C^k$-isometry then by Proposition (12.5) the
atlases ${\mathcal{A}}$ and ${\mathcal{A}}'$ are $C^k$-compatible.
Hence $\Psi$ and $\Psi'$ satisfy P1. Moreover, the isometry $f$
satisfies the relation

\begin{equation} \label{iso7}
g (p') = f^\ast_{p'} g (p).
\end{equation}

Hence from (\ref{iso6}) it follows that $g' = g$. This means that
Condition P2 is satisfied by $\Psi$ and $\Psi'$ for the same
metric $g$.\\
3. If $\Psi$ and $\Psi'$ satisfy P1 generating the same
${\mathcal{D}}$ then $f$ is a diffeomorphism by Proposition (12.5)
so that
(\ref{iso6}) is true. If $\Psi$ and $\Psi'$ satisfy P2 with $g' = g$ we find that (\ref{iso7}) holds. Hence $f$ is an isometry. \\

{\bf Corollary (12.7):} Let $\Psi$ be a partial generating
function which satisfies condition P1 and P2 of Section 7.1.2,
i.e.\ $\Psi$ generates the metric $g$ and the differential
structure ${\mathcal{D}}$, and let $f$ be a $C^k$-diffeomorphism,
$k \ge 3$. Then $f$ is an isometry exactly if

\begin{equation} \label{iso8}
\Psi (f (p), f(q)) = L (q) \cdot \Psi (p, q) + R (p, q)
\end{equation}

where $L$ is a  field of Lorentz matrices and where $d R (\cdot, q) |_{p=q} = 0$.\\
This follows directly from the Propositions (12.6) and (7.5).\\

Equation (\ref{iso8}) is the symmetry condition of $\Psi$ for a
given isometry $f$ which  we are looking for in this section. It
can be helpful to derive a special form for $\Psi$ which reduces
the very complicated generalized Einstein equation.
Similar results hold for conformal mappings, too.\\

\section{Final Remarks}
\subsection{Results}

The main point of all three parts of this treatise is the
existence of a function $\Psi$ which generates an atlas
${\mathcal{A}}$ of pre-radar charts, a metric $g$, a velocity
field $v$ and the integral curves of $v$. It is shown that also an
orientation and a time orientation is defined by $\Psi$. Finally,
the concept of isometry can be formulated directly with the help
of $\Psi$, i.e.\ without using the
metric $g$.\\
This illustrates the significance $\Psi$ has: it determines like a
"potential" almost all of the fundamental concepts considered
in GR and it is itself physically interpretable as a set of coordinate functions.\\
Since the existence of $\Psi$ guarantees the existence of a smooth
field of tetrads, it imposes restrictions on space-times. On the
other hand, by space-time theory the existence of pre-radar or
even radar charts is indispensible for space-times so that
the restrictions imposed on them by $\Psi$ are physically motivated and natural.\\
Finally, in Chapter 11 it is shown that GR can be formulated as a
scalar field theory. But the  price is doubling the independent
variables and a generalized Einstein equation which is of third
order.

\subsection{Open problems}

{\bf 13.2.1:} The main problem of any axiomatic formulation of GR,
e.g.\ of the theories $\Phi_R, \Phi^\ast_R$ and $\Phi_{sc}$, is
how to get models. For this purpose the additional conditions have
to be concreted. This can be done for $\Phi_R$ and $\Phi^\ast_R$
in the usual way, but for $\Phi_{sc}$ it is not known up to now
how to formulate a well-posed Cauchy problem for the generalized
Einstein equation together with the generalized equation of
continuity. In both cases solving Einstein's equation is the most
difficult step in obtaining models, but it is not all
one has to do. The other axioms must be satisfied, too.\\
{\bf  13.2.2:} An  open practical problem is the exploitation of
Equation (\ref{iso8}), the definition of isometry in terms of
$\Psi$. The aim is obtaining a restricted form of $\Psi$ for a
given symmetry $f$. For this purpose one needs a sufficiently
large set of representation theorems. But, little is known in this field.\\
{\bf 13.2.3:} A solution of the inverse problem described in
Section 8.1 is of physical significance because the existence of a
generating function imposes restrictions on a space-time.
Therefore it would be of great interest to find necessary
and sufficient conditions for the solution of the (non-local) inverse problem.\\
{\bf 13.2.4:} A more principal problem is the formulation of the
equations EM$_{sc}$ 3, 4 and EE$_{sc}$ in geometrical terms
without use of coordinates. It seems to me that for this goal the
product manifold $M \times M$ must be considered. Up to
now the problem is unsolved.\\
{\bf 13.2.5:} The generalized equation of continuity and the
generalized Einstein equation form a system of 11 equations for
the 6 unknown functions $\Psi, \tilde{\eta}, \tilde{\vartheta}$.
This fact seems to be a hint that there are internal dependencies
between these equations which are not known up to now. It could
also be the case that the generalized equations of continuity and
of motion together with a reduced version of the generalized
Einstein equation, e.g.
its trace, suffice to determine $\Psi, \tilde{\eta}, \tilde{\vartheta}$. \\
{\bf 13.2.6:} It is a hard task to calculate the coefficients $M$
and $K$ which occure in  Equation (\ref{prop10.8-2}) for a given
special ansatz of $\Psi$ based on  symmetries. A well adapted
computer algebra could be helpful in this field. I think that this
problem is solvable, though it is not yet solved.\\

{\bf Acknowledgement}  \\

I want to thank Mr. Gerhard Lessner for valuable discussions and
critical remarks and Mr. Wolfgang Rothfritz for correcting my
English. \\

\bibliographystyle{my-h-elsevier}

\end{document}